\newif\ifieee  
\ieeetrue
\newif\ifieeefinal
\ieeefinaltrue
\newif\iftodo       
\todofalse
\newif\iftodoshort  
\todoshortfalse
\ifieee
  \ifieeefinal
    \documentclass[a4paper,conference,twoside,twocolumn]{IEEEtran}
  \else
    \documentclass[a4paper,onecolumn,draftcls]{IEEEtran}
  \fi 
\else
  \documentclass[11pt,fleqn]{article}
\fi 
\ifieee
\else
\fi
\usepackage{amsmath}
\usepackage{amssymb}
\ifieee
  \usepackage{theorem}
\else
  \usepackage{amsthm}
\fi
\usepackage[dvips]{graphicx}
\usepackage{xspace}
\usepackage{nicefrac}
\usepackage[hyphens]{url}
\newcommand{\emptypage}%
{
  \newpage
  \vspace*{10cm}
  \pagebreak
}
\ifieee

\else

\fi
\iftodo
  \usepackage[german,english]{babel}   
  \usepackage{color}
\fi
\ifieee
  \theoremstyle{plain}
  \theoremheaderfont{\bf}
  {\theorembodyfont{\it}
  \newtheorem{theorem}{Theorem}[section] 
  \newtheorem{proposition}[theorem]{Proposition}
  \newtheorem{lemma}[theorem]{Lemma}
  \newtheorem{corollary}[theorem]{Corollary}
  \newtheorem{conclusion}[theorem]{Conclusion}
  }
  {\theorembodyfont{\rm}
  \newtheorem{definition}[theorem]{Definition}
  }
  {\theorembodyfont{\rm}
  \newtheorem{remark}[theorem]{Remark}
  \newtheorem{note}[theorem]{Note}
  \newtheorem{cit}[theorem]{Citation}
  }
\else
  \theoremstyle{plain}
    
  \newtheorem*{theorem*}{Theorem}
  
  \newtheorem*{proposition*}{Proposition}
  
  \newtheorem*{lemma*}{Lemma}
  
  \newtheorem*{corollary*}{Corollary}
  
  \newtheorem*{conclusion*}{Conclusion}
  \theoremstyle{definition}
  
  \newtheorem*{definition*}{Definition}
  \theoremstyle{remark}
  
  \newtheorem*{remark*}{Remark}
  
  \newtheorem*{note*}{Note}
  
  \newtheorem*{cit*}{Citation}

\fi 

\ifieee\else
  
\fi
\newcommand{\mapset}[2]{#1\longrightarrow #2}
\newcommand{\mapto}[2]{#1\longmapsto #2}
\usepackage{dsfont}
\newcommand{\CC}{\ensuremath{\mathds{C}}}   
\newcommand{\RR}{\ensuremath{\mathds{R}}}

\newcommand{\mdef}{:=}

\newcommand{\oneover}[1]{\frac{1}{#1}}

\newcommand{\norm}[1]{\lVert#1\rVert}

\newcommand{\abs}[1]{\lvert#1\rvert}

\newcommand{\generate}[1]{\left\langle #1 \right\rangle}
\newcommand{\smallgenerate}[1]{\langle #1 \rangle}

\newcommand{\fnorm}[1]{\lVert#1\rVert_{\scriptscriptstyle\text{F}}}
\newcommand{\bigfnorm}[1]{\left\lVert#1\right\rVert_{\scriptscriptstyle\text{F}}}

\newcommand{\Unity}{\ensuremath{\mathbf{1}}\xspace}
\newcommand{\Zero}{\ensuremath{\mathbf{0}}\xspace}

\newcommand{\ubar}[1]{\smash[b]{\underset{\rule[5pt]{0.75ex}{0.25pt}}{#1}}}

\newcommand{\textmatrix}[1]{
   \left(\begin{smallmatrix}#1\end{smallmatrix}\right)}

\DeclareMathOperator{\tr}{tr}

\DeclareMathOperator{\dimr}{\dim_{\RR}}

\newcommand{\T}{\ensuremath{T}\xspace}
\newcommand{\nt}{\ensuremath{n_{t}}\xspace}
\newcommand{\nr}{\ensuremath{n_{r}}\xspace}
\newcommand{\Gc}[1]{\ensuremath{G^{\CC}_{#1}}}  
\newcommand{\Vc}[1]{\ensuremath{V^{\CC}_{#1}}}  
\newcommand{\gnt}{\Gc{\nt,\T}}
\newcommand{\vnt}{\Vc{\nt,\T}}
\newcommand{\kbein}{\ensuremath{\textmatrix{\Unity\\\Zero}}}
\newcommand{\starred}[1]{\ensuremath{#1^*#1}}
\newcommand{\sym}{\mathrm{sym}}
\newcommand{\deltakc}{\ensuremath{\Delta}}
\newcommand{\deltauc}{\ensuremath{\underline{\Delta}}}
\newcommand{\varrhokc}{\ensuremath{\varrho}}
\newcommand{\varrhouc}{\ensuremath{\underline{\varrho}}}
\newcommand{\dkc}{\ensuremath{\mathrm{d}}\xspace}
\newcommand{\duc}{\ensuremath{\underline{\mathrm{d}}}\xspace}
\newcommand{\pkc}{\ensuremath{\mathrm{p}}\xspace}
\newcommand{\puc}{\ensuremath{\underline{\mathrm{p}}}\xspace}
\newcommand{\chkc}{\ensuremath{\mathrm{ch}}}
\newcommand{\chuc}{\ensuremath{\underline{\mathrm{ch}}}}
\newcommand{\divkc}{\ensuremath{\mathcal{D}iv}}
\newcommand{\divuc}{\ensuremath{\underline{{\cal D}iv}}}
\title{Space Time Codes from Permutation Codes}
\author{
  Oliver Henkel\\
  Fraunhofer German-Sino Lab for Mobile Communications - MCI\\
  Einsteinufer 37, 10587 Berlin, Germany\\
  Email: {\tt henkel@hhi.fraunhofer.de}
     }
\date{}
\begin{document}
\maketitle
\begin{abstract}
A new class of space time codes with high performance is presented. The
code design utilizes tailor-made permutation codes, which are known to
have large minimal distances as spherical codes. A geometric connection between
spherical and space time codes has been used to translate them into the
final space time codes. Simulations demonstrate that the performance
increases with the block lengths, a result that has been conjectured
already in previous work. Further, the connection to permutation codes
allows for moderate complex en-/decoding algorithms. 
\end{abstract}
\section{Introduction}
In MIMO (Multiple Input Multiple Output) systems space time coding schemes
have been proven to be an appropriate tool to exploit the spatial diversity 
gains. Two distinct scenarios are common, whether the channel coefficients
are known (coherent scenario) \cite{tar.ses.cal},
to the receiver or not (non-coherent scenario) \cite{hoc.mar.2}.
Prominent coherent codes are the well known Alamouti scheme \cite{ala} and
general orthogonal designs \cite{tar.jaf.cal}. A more flexible coding
scheme are the so-called linear dispersion codes. They have been introduced in
\cite{has.hoc} and were further investigated in \cite{goh.dav}. A high rate
example achieving the diversity multiplexing tradeoff is the recently
discovered Golden code \cite{bel.rek.vit}. 
Genuine non-coherent codes have been proposed in \cite{tar}, but most of
the research efforts in the literature focus on differential schemes,
introduced in \cite{hoc.swe}, since differential codes usually provide
higher data rates than comparable non differential codes. High performing
examples have been constructed in \cite{sho.has.hoc.swe},
\cite{han.ros},\cite{lia.xia},\cite{wan.wan.xia}. 
However, in both cases most research effort has been undertaken for space
time block codes with quadratic 2-by-2, resp. \nt-by-\nt\ code
matrices (\nt\ denotes the number of transmit antennas). Although linear
dispersion codes are not restricted to quadratic 
shape of the design matrices the block length is not a free design
parameter when the number of transmit antennas is held fixed (compare the
asymptotic guidelines in \cite{goh.dav}).

In contrast to that, both coherent and
(non-differential) non-coherent case are expected to benefit from coding
schemes which use the additional degrees of freedom provided by increasing
the block length \cite{hen-transinf1} (whereas \nt\ is fixed). This result
has originally been 
developed in the context of packing theory, but in \cite{hen-transinf2} its
influence on the performance on space time block codes has been pointed
out. Roughly speaking, space time code design can be considered as a
constrained sphere packing problem, where the objective (performance gain)
can be optimized in a two stage process. Step one aims to construct good
packings, step two is concerned with the maximization of the coding gain,
given a packing configuration. This method works for the coherent scenario
as well as for the non-coherent system.

The present work utilizes the proposed two stage process to construct space
time codes for both scenarios. It turns out that the performance in terms
of bit error rates of the constructed codes increases with the block
length, in accordance to what has been conjectured in
\cite{hen-transinf1}. The simulation results show, that it is possible to
beat the performance of some optimal conventional 2-by-2 schemes
considerably. 

The two optimization steps, though different in their nature, are commonly
formulated in geometric terms, according to the underlying geometric
structures of the coding spaces. While the second step is simply a suitably
defined rotation of the data (precoding in some sense), the first step
involves geometric and combinatorial aspects. The differential geometric
aspects have been already analyzed in previous publications
\cite{hen-transinf1}, \cite{hen-transinf2}, \cite{hen.wun-itg05},
\cite{hen-isit05}, and the contribution of this work has its focus on the
combinatorial part, namely the construction of appropriate spherical
permutation codes. 

Section \ref{s.model} introduces the channel model and basic definitions,
section \ref{s.design-criteria} states the code design criteria with
emphasis on the aspects which become important for the further development,
in particular subsection \ref{ss.implications-code-design} summarizes the
main points. Section \ref{s.s2stc} sketches the results of previous work,
namely the differential geometric connection between spherical packings ---
which occur e.g. in the context optimal sequence design
in CDMA systems --- and packings on the Stiefel and Grassmann manifolds, 
the appropriate coding spaces for space time block code design. 
Then in section \ref{s.perm2spherical} permutation codes 
enter the stage, since they carry naturally the interpretation as spherical
packings. The design of permutation codes yielding large packing distances
on spheres with prescribed dimension and rate requirements will be
investigated, followed in \ref{s.fulldivrot} by an analysis of the second
optimization step, i.e. the design of an appropriate rotation matrix. 
Section \ref{s.simulations} presents simulations of bit error performance
and \ref{s.conclusion} summarizes the work done so far, followed by an
outlook to further work. 
\section{Channel model and coding spaces}
\label{s.model}
Let us assume a MIMO system with \nt
transmit antennas and \nr receive antennas. The fading statistic is assumed
to obey a Rayleigh flat fading model with block length \T\ of the coherence
interval. Then we have the transmission equation
\begin{equation}\label{e.channel}
   Y = \sqrt{\rho}\, X H + N
\end{equation}
where $X$ denotes the \T-by-\nt transmit signal with normalized
expected power per time step, $H\sim\CC\mathcal{N}(\Zero,\Unity)$ is the
\nt-by-\nr circular symmetric complex normal distributed channel matrix,
$N\sim\CC\mathcal{N}(\Zero,\Unity)$ denotes the \T-by-\nr additive noise,
and $Y$ the \T-by-\nr received signal, where $\rho$ turns out to be the
SNR at each receive antenna. The symbol $\Unity$ denotes a unit matrix
throughout this work, sometimes supplemented by an index indicating the
dimension. 

Due to the work of Hochwald/Marzetta \cite{hoc.mar.1} it is reasonable from
a capacity perspective to assume the transmit signals $X$ to have (apart
from a scaling factor) unitary columns. More precisely we can write
\begin{equation}
   X=\sqrt{\frac{\T}{\nt}}\,\Phi
\end{equation}
and consider the complex Stiefel manifold
\begin{equation}\label{e.defvkn} 
   \Vc{\nt,\T} \mdef \{ \Phi\in\CC^{\T \times \nt} \,|\, \Phi^*\Phi=\Unity_{\nt} \}
\end{equation}
as the coding space ($\cdot^*$ denotes the hermitian conjugate). Thus a
space time code is considered to be a discrete 
subset $\mathcal{C}\subset\Vc{\nt,\T}$ and we define the rate $R$ of the
code by
\begin{equation}\label{e.defrate}
   R \mdef \frac{1}{\T}\log_2\abs{{\cal C}}
\end{equation}
Provided a received signal $\tilde{Y}=\sqrt{\rho\frac{\T}{\nt}}\Psi+N$ 
the maximum likelihood (ML) detection rule reads
\begin{equation}\label{e.ml-detection-kc}
   \Phi_{\text{ML}} = \arg\min_{\forall_{\Phi\in{\cal C}}}
      \bigfnorm{\tilde{Y}-\sqrt{\rho\frac{\T}{\nt}}\,\Phi H}
\end{equation}
where $\fnorm{A}=\sqrt{\tr A^*A}$ denotes the Frobenius norm. 

\subsection{Non-coherent detection}
If the receiver has no information about the fading states the detection
is called non-coherent. In this case it is shown in
\cite{hoc.mar.1,hoc.mar.2,zhe.tse} that the coding space is the complex
Grassmann manifold
\begin{equation}\label{e.defgkn} 
   \Gc{\nt,\T} \mdef \{ \generate{\Phi} \,|\, \Phi\in\Vc{\nt,\T} \}
\end{equation}
of \nt-dimensional linear complex subspaces of $\CC^{\T}$
($\generate{\Phi}$ denotes the vector space spanned by the columns of the
matrix $\Phi$). One can think of $\Phi$ representing a subspace
$\generate{\Phi}$, but for a given $\Phi\in\Vc{\nt,\T}$ all matrices
$\Phi u$ with arbitrary unitary \nt-by-\nt matrix represent the same
subspace; therefore the Grassmann manifold is really a coset space of the
Stiefel manifold and the choice of a unique representative for each
coset is not obvious in general. 
However, the maximum likelihood detection for non-coherent detection
decides on the subspace $\generate{\Phi_{\text{ML}}}$ represented by
\begin{equation}\label{e.ml-detection-uc}
   \Phi_{\text{ML}} = \arg\max_{\forall_{\generate{\Phi}\in{\cal\underline{C}}}}
      \bigfnorm{\tilde{Y}^*\Phi}
\end{equation}
given a 'received noisy subspace' $\smallgenerate{\tilde{Y}}$ represented by
$\tilde{Y}=\sqrt{\rho\frac{\T}{\nt}}\Psi+N$. Since the Frobenius norm is
unitarily invariant, the ML criterion \eqref{e.ml-detection-uc} is
independent of the chosen representatives $\Phi$ and $\Psi$, thus
\eqref{e.ml-detection-uc} provides a well defined measure of subspace
correlation. Therefore, the explicit choice of a representative $\Phi$ of 
$\smallgenerate{\Phi}\in\mathcal{\underline{C}}$ is irrelevant and we are
free to consider  
non-coherent codes $\mathcal{\underline{C}}$ as subsets of the Stiefel
manifold $\Vc{\nt,\T}$ rather than subsets of the Grassmann manifold,
thinking in terms of representatives. As a notational convention 
entities from a non-coherent context will be underlined.
\section{Space time code design criteria revisited}
\label{s.design-criteria}
\subsection{Coherent case:}
The code design aims to maximize an appropriate functional on the set of
difference symbols $\deltakc\mdef\Phi-\Psi$. Common design criteria arise
from the familiar Chernov bound for the pairwise error probability, which
has the form \cite{hoc.mar.2} 
\begin{equation}
   \chkc = \oneover{2}\left(\prod_{i=1}^{\nt} 
     \left[
       1 + \varrhokc \sigma_i^2(\deltakc)
     \right]\right)^{-\nr}
\end{equation}
where $\varrhokc\mdef \oneover{4}\rho\frac{\T}{\nt}$ and 
$\sigma(A)=(\sigma_i(A))$ generically denotes the vector of singular values
of a matrix $A$ in decreasing order. Taking this bound as the target
functional it is immediately clear that the code design does not depend on
the number of receive antennas, and the objective becomes the maximization
of the diversity functional
\begin{equation}\label{e.divkc}\begin{split}
   \divkc \mdef \prod_{i=1}^{\nt}
     \left[
       1 + \varrhokc\sigma_i^2(\deltakc)
     \right]
     = \sum_{i=0}^{\nt}s_i\varrhokc^i 
\end{split}\end{equation}
where 
$s_j\mdef
 \sym_j(\sigma_1^2(\deltakc),\dots,\sigma_{\nt}^2(\deltakc))$ 
and $\sym_j$ denotes the $j$-th elementary symmetric polynomial defined by
$\sym_j(x_1,\dots,x_{\nt})\mdef 
    \sum_{1\le i_1\le\dots\le i_j\le \nt} x_{i_1}\cdots x_{i_j}$.

The diversity contains as its first order term the receiver metric itself,
the so-called diversity sum
\begin{equation}\label{e.dkc}
   \dkc^2 \mdef s_1 = \fnorm{\deltakc}^2 
\end{equation}
as well as the diversity product as its leading term
\begin{equation}\label{e.pkc}
   \pkc^2 \mdef s_{\nt} = \det(\starred{\deltakc})
\end{equation}
\subsection{Non-coherent case:}
Following \cite{hoc.mar.2} a similar derivation applies: Defining the 
codeword difference symbol as $\deltauc\mdef\Phi^*\Psi$ the Chernov bound
now reads
\begin{equation}
   \chuc = \oneover{2}\prod_{i=1}^{\nt}
                   \left[
                     1 + \varrhouc (1-\sigma_i^2(\deltauc))
                   \right]^{-\nr}
\end{equation}
where $\varrhouc \mdef \frac{\varrhokc^2}{\varrhokc+\oneover{4}}$, and
the corresponding diversity quantities become
\begin{equation}\label{e.divuc}
   \divuc \mdef \prod_{i=1}^{\nt}
     \left[
       1 + \varrhouc(1-\sigma_i^2(\deltauc))
     \right]
     = \sum_{i=0}^{\nt}\ubar{s}_i\varrhouc^i 
\end{equation}
with
$\ubar{s}_i \mdef \sym_i\big((1-\sigma_1^2(\deltauc)),\dots,
                           (1-\sigma_{\nt}^2(\deltauc))\big)$, and
\begin{gather}
   \label{e.duc}
   \duc^2 \mdef \ubar{s}_1 = \nt-\fnorm{\deltauc}^2\\
   \label{e.puc}
   \puc^2 \mdef \ubar{s}_{\nt} = \det(\Unity-\starred{\deltauc})
\end{gather}
\subsection{Implications for the code design and known results}
\label{ss.implications-code-design}
Coherent and non-coherent diversity functions are homogeneous polynomials,
in particular a packing gain $\mapto{\dkc}{\alpha\dkc}$
(resp. $\mapto{\duc}{\alpha\duc}$), $\alpha>1$, turns out to be equivalent
to coding 
with effective power $\alpha^2\varrhokc$ (resp. $\alpha^2\varrhouc$). Thus,
the diversity sum, which has been known as a low SNR design criterion in
the literature, also scales the SNR itself, and has therefore an impact on
the higher order terms in the diversity functional, in particular onto the
diversity product. From this insight it is reasonable to consider the code
design as a constraint packing problem. This means, that the maximization
of diversity can be split up into a two-stage optimization procedure: 
\begin{enumerate}
\item 
   Find good packings in the coding spaces $\Vc{\nt,\T}$,
   $\Gc{\nt,\T}$
\item 
   Find a transformation which maps the packings into equivalent
   packings with maximal diversity product. 
\end{enumerate}
Details about the optimality
criteria in this context can be found in \cite{hen-transinf2}.

Another important point regarding packing gains is the result obtained in
\cite[Corollary IV.2]{hen-transinf1}: The achievable minimal distances
$\dkc^2$, resp. $\duc^2$ can be lower bounded by a quantity which grows
proportionally to $\frac{\T}{\nt}$, thus there is a benefit for code
designs with large block lengths and the
codes constructed in this work benefit considerably in performance as we
will see later on. 

Since the overall complexity of code design and decoding grows also with
large block lengths, in \cite[Prop. III.4]{hen-transinf2} the inequality
$\divuc\le\divkc$ has been established, which is the diversity analogue of
the information theoretic inequality $I(X;Y)\le I(X;(Y,H))$. From this one
infers immediately that any non-coherent code can be used in a coherent
scenario without performance loss. Moreover
\cite[Thm. III.5]{hen-transinf2} states, that, given a non-coherent code
$\mathcal{\underline{C}}$, the set 
$\{ \Phi u \,|\, \Phi\in\mathcal{\underline{C}}, u\in\mathcal{\bar C} \}$ 
for any \nt-by-\nt coherent code $\mathcal{\bar C}$ is actually a coherent
space time code with diversity as least as good as the diversities of
$\mathcal{\underline{C}}$ and $\mathcal{\bar C}$. This result can be
interpreted as a complexity reduction, providing two level code design and
decoding algorithms. 
\section{Space time packings from spherical codes}
\label{s.s2stc}
Let us start with the proposed first stage optimization procedure for code design,
namely the construction of packings in \vnt\ resp. \gnt\ with large minimal
distance. A comprehensive standard source on the general sphere packing
problem in Euclidean space is \cite{con.slo}. Unfortunately the methods in
\cite{con.slo} rely on the symmetry group of Euclidean space and do not
apply to our situation, where the coding spaces are non-flat and the
distance metric is nonlinear. Although \cite{con.har.slo} considers
Grassmannian packings, it applies to the real Grassmannian manifold 
only. Some genuine complex Grassmannian
packings have been constructed numerically in
\cite{agr.ric.urb},\cite{tropp}, and \cite{goh.dav.2} but numerical
optimization techniques are computational complex and give only little
insight into the construction mechanisms nor do they possess any
algebraic structure.  

Therefore it would be desirable to find simple model spaces, where
\emph{structured} packings can be constructed and then transformed into packings on the
complex Stiefel and Grassmann manifolds. On the one hand this model space must
possess a large symmetry group such that some structured packing algorithm may
be developed. On the other hand it must be 'similar' to the Stiefel
and Grassmann manifold in order to construct a mapping which
approximately preserves (minimal) distances. 
In this paper such a model space with corresponding mapping will be
presented utilizing the 
\emph{homogeneous structure} of the coding spaces (compare \cite{gal.hul.laf} for
a general introduction to homogeneous spaces or \cite{ede.ari.smi} for the
homogeneous structure of the (real) Stiefel and Grassmann manifolds). In
particular the (complex) Stiefel manifold $\vnt$
is diffeomorphic to a coset space with respect to the unitary group
$U(\T)$ of \T-by-\T unitary matrices:
\begin{equation}
   \vnt \cong
   U(\T) \left/ \textmatrix{
       \Unity & \Zero\\
       \Zero & U(\T-\nt)
     }\right.
\end{equation}
whereas $\cong$ means 'diffeomorphic to'. This fact is due to the symmetry
action $\mapto{\Phi}{\textmatrix{\Unity & \Zero\\ \Zero & U(\T-\nt)}\Phi}$
leaving \kbein\ fixed. 
Similarly for the (complex) Grassmann manifold $\gnt$ of $\nt$ dimensional
subspaces $\generate{\Phi}$ of $\CC^{\T}$: 
Since $\mapto{\Phi}{\generate{\Phi}}$ is a projection invariant under all
\nt-by-\nt\ unitary basis transformations we obtain the coset representation
\begin{equation}
   \gnt \cong U(\T) \left/ \textmatrix{
       U(\nt)   & \Zero\\
       \Zero & U(\T-\nt)
     }\right.
\end{equation}
Homogeneity (or coset structure) means, that any two points can be mapped
isometrically into each other, in particular all distance relations are
uniquely determined with respect to an arbitrarily chosen reference point
(e.g. $\kbein$, resp. $\generate{\kbein}$). We will see that homogeneity
provides the required 'similarity' mentioned above. Let us define $D$ by 
$D =\dimr\vnt=\nt(2\T-\nt)$ resp. $D =\dimr\gnt=2\nt(\T-\nt)$.
The $D$ dimensional sphere 
$S^D\mdef \{x\in\RR^{D+1} \,|\, \norm{x}=1 \}\subset\RR^{D+1}$ 
is also homogeneous,
since it has the coset representation 
\begin{equation}
   S^D = V^{\RR}_{1,D+1}
    \cong O(D+1)\left/ \textmatrix{
       1 & \Zero\\
       \Zero & O(D)
     }\right.
\end{equation}
where $O(D)$ denotes the set of $D$-by-$D$ orthogonal matrices. 
The sphere is highly symmetric and 'similar' to our coding spaces, since
in \cite{hen-transinf1} a relation between packing densities of the coding
spaces and $S^D$ has been
established, and in \cite{hen.wun-itg05,hen-isit05} a corresponding
mapping of packings $\mapset{S^D}{\vnt}$,
resp. $\mapset{S^D}{\gnt}$\footnote{\label{foot.gnt-rateloss}
  Actually the mapping is appropriately defined on the upper (or lower)
  hemisphere of $S^D$ only. This is due to the projective nature of \gnt
  such that antipodal points on the sphere will be identified under this
  mapping. 
} 
has been defined,
utilizing the homogeneous coset structure. 
Due to the analysis in \cite{hen-transinf1} this mapping is distance
preserving up to a positive scaling factor. 
In summary, spherical codes can be transformed into space time codes with
controlled distance loss. Moreover the theory of spherical packings
(i.e. packings of spherical caps on $S^D$) is already an item of current
research, see
e.g. \cite{ham}, \cite{hea.str.pau}. Nevertheless, here another spherical
packing algorithm will be presented to obtain 
structured and at the same time full rate spherical packings. However, 
in the space frequency context of MIMO-OFDM systems spherical packings
based on lattice constructions have already been investigated 
\cite{hen.wun-itg05,hen-isit05}. 
\section{Spherical packings from permutation codes}
\label{s.perm2spherical}
A more flexible algebraic tool than lattices to produce spherical
packings are groups, i.e. finite subgroups of the orthogonal group. The
idea behind it is to take some initial ($D+1$) dimensional vector of
unit norm (s.t. it can be considered as a point on the $D$ dimensional
sphere $S^D$). Then let the finite subgroup $G$ act on the initial vector
$x$ and
the outcome is a spherical packing whose constellation size equals the
order of $G$. The optimization procedure to maximize the packing distance
involves the choice of the group $G$ itself and the choice of the initial
vector. The packings generated by such a procedure are called geometrically
uniform and have been considered recently in a frame theoretic context
\cite{eld.boe} (see \cite{str.hea} for an introduction to frame theory in
communications). 

In a broader context the set of vectors (input sequences) obtained as
orbits of (a subset of) 
$G$ of some initial vector is called a group code for the Gaussian
channel. This class of codes comprises many signal sets that are used in
practice, e.g. linear binary codes. In the special case $G$ consisting of
$(D+1)$-by-$(D+1)$ matrix representations of permutations, the resulting
group code is called permutation modulation \cite{ingemarsson}. Note that
in practice only subgroups of the permutation group will be of interest,
otherwise the huge number of $D!$ permutations generate permutation
modulations no practical device can handle.

The corresponding spherical packings will be the starting point for the 
following analysis. In \cite{ingemarsson} an optimization procedure similar
to a Lagrangian method is
presented, which solves for the initial vector whose generated permutation
modulation has largest minimal distance under the action of a fixed
permutation subgroup. The size of the subgroup is specified in terms of the
initial vector with appropriate repetitions of its components
\begin{equation}\label{e.initialx}
   x=(\mu_1^{(m_1)},\dots,\mu_k^{(m_k)})
\end{equation} 
where
$\mu_i^{(m_i)}$ denotes $\mu_i$ repeated $m_i$ times. Although the analysis in
\cite{ingemarsson} does not provide a complete solution (no solution for the
'Lagrangian' parameters has been given), the method reveals some structure of
the optimal initial vector: The entries $\mu_i$ are symmetrically arranged
around zero and the corresponding weights $m_i=\lfloor
e^{-(\eta+\mu_i^2)/\lambda}\rceil$ are determined according to 
some discrete Gaussian distribution involving the 'Lagrangian'
parameters $(\eta,\lambda)$ \cite[Sec. IV]{ingemarsson}. Plugging this into
the constraint equation of the 'Lagrangian' analysis yields, using Maple,
complete 
solutions. Unfortunately due to the integer constraint on the $m_i$
solutions are possible only for carefully selected parameters. The typical
spherical dimensions $D$ occurring here do not permit solutions with small
enough rates. Therefore another strategy has been chosen.

Inspection of the initial solution vectors with lowest possible rate, such
that the 'Lagrangian' functional provides a solution, revealed that there
are only a few possible alternatives for the choice of $x$, namely $x$ is
characterized by a large amount of zero components and only a few non-zero
ones. The more distinct components in $x$, the larger the set of distinct
permutations (high rate), and the smaller the final minimal
distance. Therefore for prescribed dimension and rate the initial
vector $x$ with largest possible number of zero-components has been
chosen, such that the rate requirement is satisfied.

Having found an appropriate initial vector the problem of carefully
selecting the corresponding permutations remains. Given $x\in\RR^{D+1}$ of
the form \eqref{e.initialx} the corresponding number of distinct permuted
versions is (in multi index notation with respect to the vector
$m=(m_1,\dots,m_k)$) 
\begin{equation}
   \label{e.noMultiPerm}
   M \mdef \binom{\abs{m}}{m!}
     = \frac{(\sum_i m_i)!}{m_1!\dots m_k!}
\end{equation}
Given a prescribed space time code rate $R$, the corresponding rate of the spherical
code is $r\mdef \frac{\T}{D+1} R$ and the required number of permutations is
given as $N=\lceil 2^{(D+1)r} \rceil$, where we have chosen the initial
vector $x$ (resp. the vector $m$) such that $N\le M$ holds.
Then the task is, to select $N$ out of the $M$ distinct permutations of the
multiset\footnote{the term multiset denotes a set with repeated elements}
$x$ such that the resulting packing has large minimal distance. Taking the
number of transpositions required to transform a permutation $p$ into
another permutation $q$ as a distance measure between $p$ and $q$, the
objective is to select $N$ out of $M$ multiset permutations with large
pairwise distance. In contrast to ordinary permutations the structure of
multiset permutations is more complicated, and there seems to be no ranking
algorithm available. Nevertheless all multiset permutations can be listed
in Gray code order, which is the appropriate ordering with respect to the
permutation distance just defined. The algorithm can be obtained as a short
C program from the Combinatorial Object Server\footnote{Programmer: Frank
  Ruskey / Joe Sawada\\\url{http://www.theory.csc.uvic.ca/~cos/inf/mult/Multiset.html}}. 
Then, taking each $\lfloor \frac{M}{N} \rfloor$'s multiset permutation
produced by this algorithm does the job and we end up with the desired
spherical packing with large minimal distance, corresponding to the
specified rate. 
\section{Full diversity rotation}
\label{s.fulldivrot}
\begin{figure}[htb]
\begin{center}
   \includegraphics[width=.9\linewidth,height=5cm]{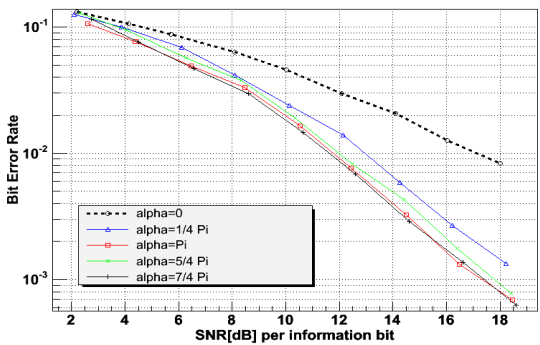}
\end{center}
   \caption{\label{fig.alphaseries}
     Performance of $R=.5, 8\times2$ space time codes coming from the same 
     spherical code, but precoded with different rotation angles
   }
\end{figure}

Let us now come the the second stage of diversity optimization in the sense
described in \ref{ss.implications-code-design}, namely to define a
distance preserving mapping which transforms the space time packings into
an equivalent packing with maximum diversity product. To this end we precode the
space time code symbols by performing a rotation on the spherical code as
follows. As the axis of rotation we choose the 'diagonal'
$e=(1,\dots,1)\in\RR^{D+1}$. Define a unitary $(D+1)$-by-$(D+1)$ matrix $W_e$ by
prescribing its first row to be $e/\sqrt{D+1}$ and for $j=2,\dots,D+1$ its
$j$th row to be $(1^{(j-1)},-j(j-1),0^{(D+1-j)})/\sqrt{j(j-1)}$. Clearly
$e=e_1 W_e$ holds with 
$e_1=(1,0,\dots,0)$, thus $e_1=e W_e^t$, where the superscript $t$
denotes transposition. Suppose we already had defined a rotation matrix $R_1$
with $e_1$ as its axis, then we obtain the same rotation about the axis $e$
as $R\mdef W_e^t R_1 W_e$. The rotation $R_1$ is constructed easily: 
Set $\Zero=(0^{(D)})$, then 
$R_1=\textmatrix{
              1     & \Zero\\
              \Zero^t & \exp(\alpha X)
     }
$
performs a rotation about $\alpha$ degrees about the axis $e_1$, where
$X$ being the antisymmetric $D \times D$ matrix with ones on its
upper triangular part (which uniformly weights the available degrees of
freedom). 
Figure \ref{fig.alphaseries} demonstrates the effect of rotation for some
values of $\alpha$ on the performance of a sample non-coherent $8\times2$
code of rate $\nicefrac{1}{2}$. Note that without rotation ($\alpha=0$,
thick dashed line) the
code does not achieve full diversity order. Trying some values for $\alpha$
reveals some oscillatory behavior of the coding gain (i.e. the value of
the diversity). It turns out that for non-coherent codes
$\alpha=\frac{7}{4}\pi$ is a good choice, while for coherent codes
$\alpha=\pi$ yields good results. If a non-coherent code will be used in
the coherent scenario by composing it with some small coherent code
(compare \ref{ss.implications-code-design}), the angle
$\alpha=\frac{7}{4}\pi$ remains a good choice. 
\section{Simulation results}
\label{s.simulations}
All simulations have been performed in a scenario with $\nt=2$ transmit
antennas and $\nr=1$ receive antennas with maximum likelihood decoding.
\begin{figure}[htb]
\begin{center}
   \includegraphics[width=\linewidth,height=6cm]{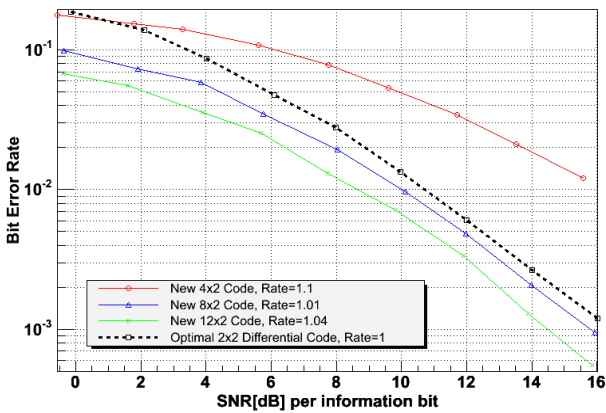}
\end{center}
   \caption{\label{fig.simuc}
   Non-coherent performance gain with increasing block length, compared to
   the optimal 2-by-2 differential code}
\end{figure}
Figure \ref{fig.simuc} displays the bit error performance of a series of
two-stage-optimized non-coherent codes with rate approximately one and
block lengths varying from 4 to 12 (continuous lines). The corresponding
initial vectors (of dimension $D+1$) and the number of chosen
multiset permutations are 
$x=(0^{(7)},1^{(2)})/\sqrt{2}$, $N=32$;
$x=(-1,0^{(23)},1)/\sqrt{2}$, $N=512$;
$x=(0^{(38)},1^{(3)})/\sqrt{3}$, $N=8192$, respectively.
The rotation angle is $\alpha=\frac{7}{4}\pi$ and the final space time
code is then given as the image of the map $\mapset{S^D}{\gnt}$ 
(compare section \ref{s.s2stc}), where now (for $\nt=2$ fix)
$D=8,24,40$ for $T=4,8,12$ respectively. Note that the cardinality of
the final space time codes differs from the corresponding spherical code
cardinality due to the restriction to one hemisphere of $S^D$, compare
footnote~\ref{foot.gnt-rateloss} in section \ref{s.s2stc} (e.g. the
spherical code of cardinality $N=32$ shrunk to a space time code of
cardinality $21$ only, thus $R\approx 1.1$).
The simulation shows
that the bit error performance increases with the block length in perfect
conformity with the result of earlier work \cite{hen-transinf1}, mentioned
in \ref{ss.implications-code-design}. Moreover \cite{lia.xia} presented a
non-coherent 2-by-2 differential code with optimal diversity sum and diversity
product. The performance of this optimal 2-by-2 code is also shown in
fig.~\ref{fig.simuc} (thick dashed line). The comparison reveals that the
additional degrees of freedom provided by the larger block lengths of the
new codes based on permutation codes result in an approximately 2dB
performance gain over the 2-by-2 differential code \cite{lia.xia}. 
Note that the non-coherent codes constructed here are not based on a
differential transmission scheme. Thus the achieved performance gain over
one of the best known differential schemes justifies the research effort
for non-differential schemes.

\begin{figure}[htb]
\begin{center}
   \includegraphics[width=\linewidth,height=6cm]{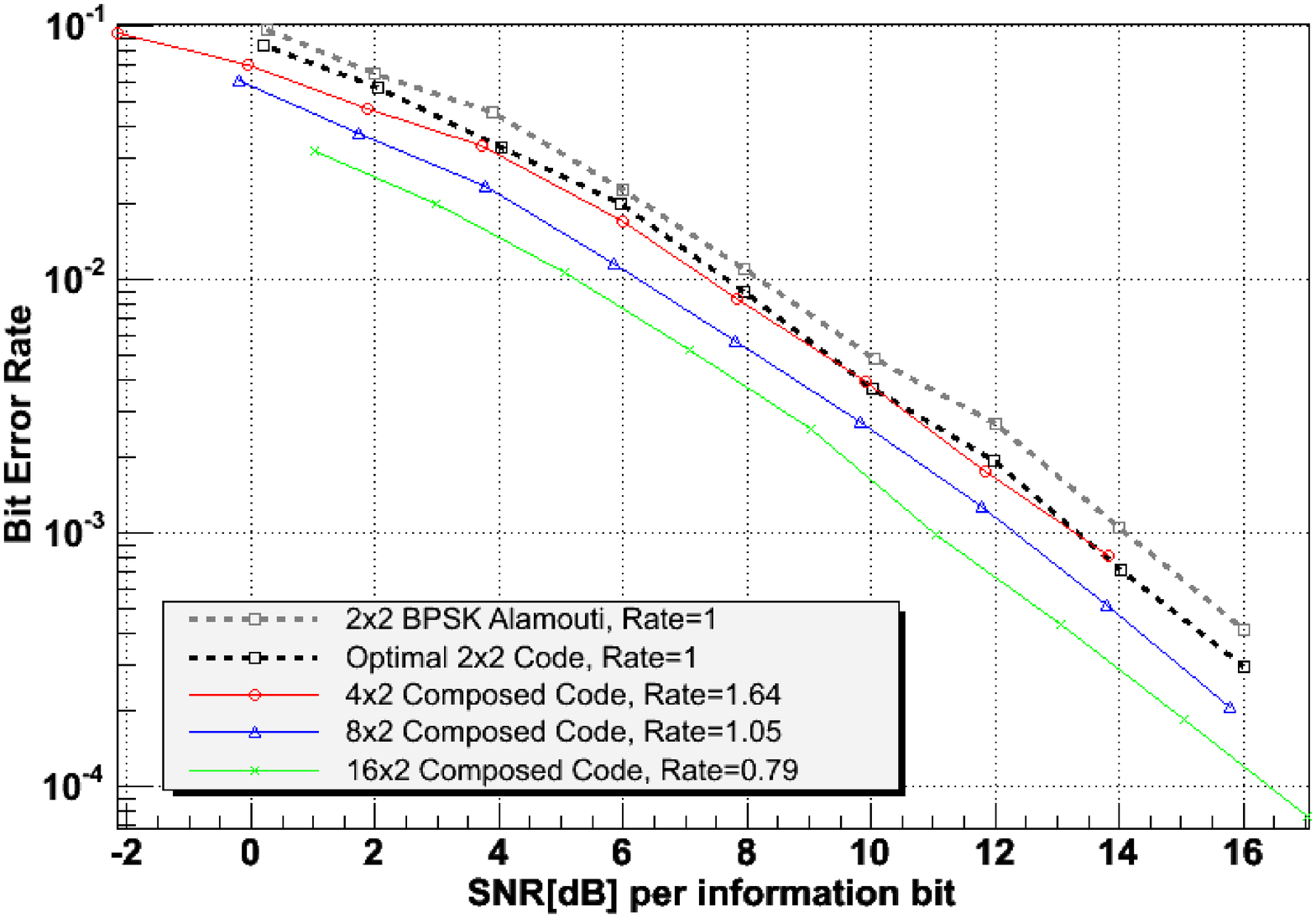}
\end{center}
   \caption{\label{fig.simkc}
   Coherent performance gain with increasing block length, compared to the
   well known BPSK Alamouti scheme}
\end{figure}
Figure \ref{fig.simkc} displays the bit error performance of a series of
two-stage-optimized composed coherent codes with rates ranging from 1.64 to
0.79 and block lengths $T=4,8,16$ (continuous lines). They have been 
composed from a series of non-coherent codes and a QPSK Alamouti
scheme \cite{ala}. The non-coherent codes come from corresponding spherical
codes of size $N=8,32,512$ (where again some spherical code points have
been removed due to the restriction to only one hemisphere) and dimension $D=8,24,56$.
Again the bit error performance increases
with the block length and comparing the rate 1.05 8-by-2 code
with the 2-by-2 BPSK Alamouti code (thick dashed gray line in fig.~\ref{fig.simkc})
shows a performance gain of approximately 2dB. Of course the new codes
suffer from a considerable higher decoding complexity compared
with the Alamouti scheme, thus there is a tradeoff between performance and
signal processing. A more fair comparison incorporating some additional
signal processing may be represented by the thick dashed black
line in fig.~\ref{fig.simkc}. It shows the performance of a 2-by-2 code with
optimal diversity sum and diversity product, which is in fact identical to
the optimal non-coherent 2-by-2 differential code
\cite{lia.xia}\footnote{This is due to the fact, that the code design
  criteria for differential space time codes coincide with the design
  criteria for quadratic coherent space time code matrices. Therefore
  optimal differential codes yield optimal coherent codes}. 
This code performs about 1dB better than the Alamouti scheme but compared
with the new codes we still obtain a performance gain of approximately 1dB
of the new 8-by-2 code over the optimal 2-by-2 code.
\section{Conclusions and future work}
\label{s.conclusion}
A new class of space time codes based on spherical permutation codes has
been presented. It has been demonstrated that the additional degrees of
freedom provided by larger block lengths help to achieve better performance
and even beat the bit error performance of 2-by-2 diversity-optimal schemes. 
The presented construction applies both to coherent and non-coherent code
design with a two-stage optimization process which reduces the design
complexity by geometrical insights affording algebraic structures.

The inherent design complexity of coherent codes with large block lengths
can be further compensated in part by reduction to the design of
non-coherent codes, supplemented by small coherent codes. 
The non-coherent code design in turn is not 
based on any differential scheme but on the packing theory of the Grassmann
manifold.

However, the presented construction scheme, in particular the use of
permutation codes will be investigated further, in order to obtain low
complex decoding algorithms in the future.
\section*{Acknowledgment}
I want to express my gratitude to my colleague Gerhard Wunder for pointing
out to me the reference \cite{ingemarsson}, which served as a fruitful
starting point of this work.
%
%
%
%
%
%
%
%
%
%
%
%
\bibliography{%
/home/henkel/texstuff/bib/ieee/IEEEabrv,%
/home/henkel/texstuff/bib/refmci,%
/home/henkel/texstuff/bib/myrefs} 
\end{document}